\documentstyle[axodraw,a4]{article}

\newcommand{\be}{\begin{equation}}
\newcommand{\ee}{\end{equation}}
\newcommand{\ba}{\begin{eqnarray}}
\newcommand{\ea}{\end{eqnarray}}

\newcommand{\slp}{\not \! p}

\newcommand{\tr}{ {\rm tr} }

\begin{document}

\begin{titlepage} 

\begin{flushright}
OUTP-98-13P \\
hep-th/9802146
\end{flushright}

\begin{centering} 
\vspace{2cm} 

{\bf Composite Operator Effective Potential Approach to QED$_3$}

\vspace{2cm} 

{\bf A. Campbell--Smith}

\vspace{1cm} 

University Of Oxford, Department of Physics, 
Theoretical Physics,  
1 Keble Road, OXFORD, OX1 3NP, U.K.

\vspace{2cm} 

{\bf Abstract} 
\vspace{1cm} 

\end{centering}

The composite operator effective potential is compared with the
conventional Dyson--Schwinger method as a calculational tool for
(2+1)-dimensional quantum electrodynamics.  It is found that when the 
fermion propagator {\sl ansatz} is put directly into the effective
potential, it reproduces exactly the usual gap equations derived in
the Dyson--Schwinger approach.

\end{titlepage} 

\newpage
\section{Introduction.}

Studies of $N$-flavour Quantum Electrodynamics in three dimensions
(QED$_3$) have long been made in order to learn about dynamical mass
generation (and hence chiral symmetry breaking) in a relatively simple
context \cite{pis,ap1,ap2,pen}.  Initially the hope was to gain
insight into dynamical symmetry breaking in other asymptotically free
theories, in particular Yang--Mills theories in four
dimensions. Subsequently there has been interest in QED$_3$ as a model
for high-T$_{\rm c}$ superconductors \cite{dma}.  Here the broken,
massive phase of QED$_3$ gives a mechanism for spontaneously breaking
the $U(1)$ gauge symmetry which is necessary for superconductivity,
and deviations from the trivial infra red fixed point could be
responsible for the observed non-Fermi liquid behaviour in the normal
phase of these materials \cite{ama}.

Considering $N$ fermion flavours allows one to perform a large $N$
expansion and keep only dominant terms in this limit.  This is a
convenient calculational tool for it admits a form for the full gauge
boson propagator (namely that with massless fermion loops summed
\cite{pis}) without requiring that the Dyson--Schwinger equation for
the propagator be solved.  There is a further significance of working
with $N$ flavours in the possibility that there exists a critical
number of flavours \cite{ap1,amm} above which no symmetry breaking
occurs.  Much of the evidence for the existence of a critical flavour
number lies in numerical work and the analytical studies that have
been made require great simplifications to be tractable; indeed there
is evidence for no critical behaviour in the flavour number (see
e.g. \cite{pen}).

Non-perturbative treatments of QED$_3$ are normally based on solving
systems of Dyson--Schwinger equations.  To make the systems soluble
the equations for the gauge boson self energy are dropped in favour of
the propagator resummed in large $N$; the equations for the vertex are
replaced by an {\sl ansatz} discussed further below.  This leaves only
the Dyson--Schwinger equation for the fermion propagator to be solved;
the approach taken is to substitute a (rather general) {\sl ansatz}
for the propagator into the equations and then solve for the functions
appearing in the {\sl ansatz}.

The system of Dyson--Schwinger equations for a theory can in principle
be derived from a composite operator effective action \cite{cjt} via a
well-defined minimization technique.  It is the purpose of this paper
to consider whether the same physics is found if one puts the {\sl
ansatz} directly into the effective action.  The same minimization
technique can then be used to derive the ``gap equations'' for the
functions in the {\sl ansatz} directly from the effective action, and
these can then be compared with the conventional Dyson--Schwinger
results.  The system of Dyson--Schwinger equations used is heavily
truncated and the effective action is computed to two loop level only;
it is therefore not immediately obvious that the results of the two
computations will coincide.  Indeed in general the calculations may
give different answers (e.g. \cite{gac}), and comparing the methods
can allow one to estimate the limitations of the approximations used.
Since the approximations described above are used almost universally
in work on QED$_3$ it is important to compare the two methods within
the those limits.  It will be shown that the composite operator
effective action method gives exactly the usual (Dyson--Schwinger) gap
equations, after simplification of the ``raw'' equations which come
directly from the minimization procedure.  The amount of computational
effort required is rather surprising, and may have important
consequences for future work at finite temperature; it is probable
that finite temperature Dyson--Schwinger approaches would be more
economical.

This paper is organized as follows: in section \ref{effac}, the basic
formalism for composite operator effective actions is reviewed, rewritten
for fermionic fields; section \ref{cjtvsds} contains a comparison of the
effective action and Dyson--Schwinger equation approaches and in section
\ref{effgaps} are the details of the effective action calculation.  Conclusions
are appended to these sections.

\section{The Composite Operator Effective Action.} \label{effac}

The composite operator effective action \cite{cjt} is a generalization
of the conventional (quantum) effective action; for fermionic fields it depends
not only on $\psi(x)$ and $\overline{\psi}(x)$ --- possible expectation
values of the quantum fields $\Psi(x)$ and $\overline{\Psi}(x)$ --- but
also on $S(x,y)$, a possible expectation value of the time ordered
product $T\{\overline{\Psi}(x)\Psi(y)\}$.  Physical solutions require that
the effective action be stationary with respect to variations in the
expectation values:
\ba \label{statvars}
\frac{\delta \Gamma[\psi,\overline{\psi},S]}{\delta \psi(x)} &=& 0, \nonumber \\
\frac{\delta \Gamma[\psi,\overline{\psi},S]}{\delta
\overline{\psi}(x)} &=& 0, \nonumber \\
\frac{\delta \Gamma[\psi,\overline{\psi},S]}{\delta S(x,y)} &=& 0.
\ea
For this reason the formalism is especially suitable for studying
dynamical symmetry breaking, for although the first two of equations
(\ref{statvars}) may only have the symmetric solution
$\psi(x)=\overline{\psi}(x)=0$, symmetry breaking solutions may exist
for the last equation.

First we define the generating functional
$Z[\eta,\overline{\eta},\Lambda]$ ($ \int_x =\int d^3 x$):
\ba
Z[\eta,\overline{\eta},\Lambda] &=& \exp\{ i
W[\eta,\overline{\eta},\Lambda]\} \nonumber \\
&=& \int D\Psi D\overline{\Psi} \exp \left\{ i \left[
I[\Psi,\overline{\Psi}] + \int_x \left( \overline{\eta}(x) \Psi(x) +
\overline{\Psi}(x) \eta(x) \right) \right. \right.\nonumber \\
&+& \left.\left. \int_x \int_y \overline{\Psi}(x)
\Lambda(x,y) \Psi(y) \right] \right\},
\ea
where $\{\eta,\overline{\eta}\}$ and $\Lambda$ are fermionic and
bosonic source currents respectively.  The $\Psi$ and
$\overline{\Psi}$ integrations are functional, and
$I[\Psi,\overline{\Psi}]$ is the classical effective action.  The
composite operator effective action is the double Legendre transform
of $W[\eta,\overline{\eta},\Lambda]$.  Defining
\ba
\frac{\delta W[\eta,\overline{\eta},\Lambda]}{\delta \eta(x)} &=& -
\overline{\psi}(x) \, ,\nonumber \\
\frac{\delta W[\eta,\overline{\eta},\Lambda]}{\delta
\overline{\eta}(x)} &=& \psi(x) \, ,\nonumber \\
\frac{\delta W[\eta,\overline{\eta},\Lambda]}{\delta \Lambda(x,y)} &=&
\overline{\psi}(x)\psi(y) + S(x,y),
\ea
one can eliminate $\eta$, $\overline{\eta}$
and $\Lambda$ in favour of $\psi$, $\overline{\psi}$ and $S$ in the
Legendre transform:
\ba
\Gamma[\psi,\overline{\psi},S] &=& W[\eta,\overline{\eta},\Lambda] -
\int_x \left( \overline{\psi}(x) \eta(x) + \overline{\eta}(x) \psi(x)
\right) \nonumber \\
&-& \int_x \int_y \overline{\psi}(x) \Lambda(x,y) \psi(y) -
\int_x \int_y S(x,y) \Lambda(y,x)  .
\ea
It is then easy to show that
\ba \label{vars}
\frac{\delta \Gamma[\psi,\overline{\psi},S]}{\delta \psi(x)} &=&
\overline{\eta}(x) + \int_y \overline{\psi}(y) \Lambda(y,x)  ,
\nonumber \\
\frac{\delta \Gamma[\psi,\overline{\psi},S]}{\delta
\overline{\psi}(x)} &=& -\eta(x) - \int_y \Lambda(x,y) \psi(y)  ,
\nonumber \\
\frac{\delta \Gamma[\psi,\overline{\psi},S]}{\delta S(x,y)} &=&
-\Lambda(x,y)  .
\ea
Physical processes correspond to vanishing sources, so equations
(\ref{vars}) provide a derivation of the stationary requirements,
equations (\ref{statvars}).

Since we will be interested only in translation-invariant solutions we
can set $\psi$ and $\overline{\psi}$ to constants and take $S(x,y)$ to
be a function only of the separation $x-y$.  This leads to a
generalization of the effective potential \cite{cjt}, defined by
\be
V[\psi,\overline{\psi},S] \int_x = - \Gamma[\psi,\overline{\psi},S].
\ee
Then a series expansion for $V[\psi,\overline{\psi},S]$ is relatively
easy to find \cite{cjt}; the specific expansion for QED$_3$ is given
in section \ref{cjtvsds}.

Composite operator effective actions depending on possible expectation
values of irreducible three-point, four-point etc. functions can be
defined simply by taking further Legendre transforms, however the
``bilocal'' effective action defined above is sufficient to study
QED$_3$.

\section{The Effective Potential and Dyson--Schwinger Equations in
QED$_3$.} \label{cjtvsds}

The Lagrangian density for $N$-flavour QED$_3$ is as follows:
\be
{\cal L} = -\frac{1}{4} F_{\mu\nu}^2+
i\, \sum_{k=1}^{N}\overline{\Psi}_k \gamma_{\mu} {\cal D}_{\mu} \Psi_k
+ {\cal L}_{\rm GF},
\ee
where $F_{\mu\nu}$ and ${\cal D}_{\mu}$ are the usual $U(1)$ curvature
tensor and covariant derivative respectively; the fermions are
represented by four-component Dirac spinors $\Psi_k$ so that the
theory has a chiral symmetry and can exhibit mass generation
\cite{ap1}.  The gauge fixing term is not written explicitly:
throughout the calculation we shall work in Landau gauge.  The series
expansion for the composite operator effective potential in momentum
space is then \cite{cjt} ($ \int_k =\int d^3 k/(2\pi)^3$):
\ba \label{effpot}
V [S,D] & = & i\,N \int_p \tr \left\{ \ln [S^{-1}(p) S_0(p)] +
S_0^{-1}(p) S(p) -1 \right\} \nonumber \\
& - & \frac{i}{2} \int_p \tr \left\{ \ln
[D^{-1}(p) D_0(p)] + D_0^{-1}(p) D(p) -1 \right\} \nonumber \\
& + & V_2 [S,D]  .
\ea

The functions $S$ and $D$ are candidate full non-perturbative
propagators for the fermions and gauge fields respectively, and the
subscripts ${}_0$ denote their bare counterparts.  The effective
potential is minimized with respect to (functional) variations in $S$
and $D$ to determine the physical non-perturbative propagators for the
theory, as described in section \ref{effac}.  $-V_2 [S,D]$ is the sum
of all two-particle irreducible vacuum graphs with the propagators set
equal to $S$ and $D$ and with bare (not dressed) vertices.  To
consider more general vertices one would have to got to a ``trilocal''
effective action (i.e. three Legendre transforms), and require that
this new object be stationary also with respect to variations in a
vertex {\sl ansatz}.  

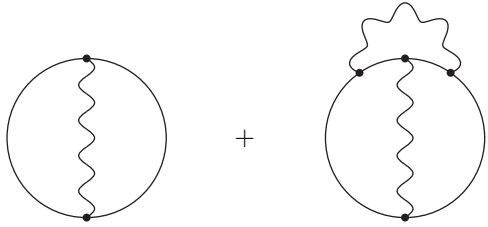
\begin{figure}
\begin{center}
\begin{picture}(200,100)(0,0)
\CArc(40,40)(30,0,360)
\Photon(40,70)(40,10){3}{4.5}
\Vertex(40,70){1.5}
\Vertex(40,10){1.5}
\Text(100,40)[]{$+$}
\CArc(160,40)(30,0,360)
\Photon(160,70)(160,10){3}{4.5}
\PhotonArc(160,70)(18,342.54,197.46){3}{4.5}
\Vertex(177.17,64.6){1.5}
\Vertex(142.83,64.6){1.5}
\Vertex(160,70){1.5}
\Vertex(160,10){1.5}
\end{picture}
\caption{Two-particle irreducible graphs contributing to $V_2$, up to
three loop level.\label{2pi}}
\end{center}
\end{figure}

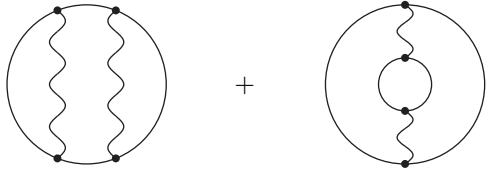
\begin{figure}
\begin{center}
\begin{picture}(200,80)(0,0)
\CArc(40,40)(30,0,360)
\Photon(29,67.91)(29,12.09){3}{3.5}
\Photon(51,67.91)(51,12.09){-3}{3.5}
\Vertex(29,67.91){1.5}
\Vertex(29,12.09){1.5}
\Vertex(51,67.91){1.5}
\Vertex(51,12.09){1.5}
\Text(100,40)[]{$+$}
\CArc(160,40)(30,0,360)
\CArc(160,40)(10,0,360)
\Photon(160,70)(160,50){3}{1.5}
\Photon(160,30)(160,10){3}{1.5}
\Vertex(160,70){1.5}
\Vertex(160,50){1.5}
\Vertex(160,30){1.5}
\Vertex(160,10){1.5}
\end{picture}
\caption{Two-particle reducible graphs not contributing to $V_2$ (to
three loop level), but which would contribute to the one-particle
irreducible effective action.\label{1pi}}
\end{center}
\end{figure}

The graphs contributing to $V_2[S,D]$ up to three loop level are
depicted in figure \ref{2pi}.  There are also two-particle reducible
graphs which would have to be considered in an ordinary (one-particle
irreducible) effective action calculation but which do not contribute
to $V_2[S,D]$; these are depicted up to three loop level in figure
\ref{1pi}.

We will truncate the series at the two loop level, as this will be
sufficient for comparison with the usual Dyson--Schwinger method for
QED$_3$.  The two-loop term in $V_2$ follows:
\be
V_2 [S,D]  =  \frac{e^2 N}{2} \int_p \int_k \tr \left\{ \gamma_\mu
S(p)\Gamma_\nu S(p+k) D_{\mu\nu} (k) \right\}.
\ee
For the purposes of comparison with the Dyson--Schwinger method we
keep the ``general'' vertex $\Gamma$; when we turn to the effective
potential calculation proper in section \ref{effgaps} this will be
reset to the bare vertex.  On demanding that $V [S,D]$ be stationary
with respect to variations in $S$ and $D$ the usual Dyson--Schwinger
equations are obtained:
\be \label{dss}
S^{-1}(p) = S_0^{-1}(p) + e^2 \int_k \, \left\{S(k) \gamma_\mu D_{\mu\nu} (q-k)
\Gamma_\nu \right\};
\ee
\be \label{dsd}
D_{\mu\nu}^{-1}(p) = {D^{-1}_0}_{\mu\nu}(p) - e^2 N \, \int_k  \, \left\{ \gamma_\mu
S(k) \Gamma_\nu S(k+p) \right\}.
\ee
Now these equations can be solved for $S$ and $D$ (with some {\sl
ansatz} or a further Dyson--Schwinger equation for the vertex) to give
the full non-perturbative propagators which minimize the effective
potential.  Since the interest is in possible dynamical mass
generation in the fermion sector we use an {\sl ansatz} for the
fermion propagator and solve for the functions appearing therein:
\be \label{anz}
S^{-1}(p) = -i \left(A(p)\slp + B(p) \right).
\ee
Here $A$ and $B$ are both assumed to be scalar functions, so that this
{\sl ansatz} is rather general in that it only omits possible
parity-breaking mass terms.

The gauge propagator is parameterized by the vacuum polarization, and
we approximate this function by the leading order truncation in $1/N$ \cite{pis}:
\be \label{gauge}
D_{\mu\nu}(p) = - \frac{i}{p^2}\frac{1}{1+\alpha /p}
\left( \delta_{\mu\nu} - \frac{p_\mu p_\nu}{p^2} \right);
\ee
$$\alpha = \frac{e^2 N}{8}  .$$

The only requirement put on the vertex function is that it is
dependent on the magnitudes alone of the momenta at the vertex, but it
must also be consistent with the Ward--Takahashi identity \cite{mar}:
\be \label{vertex}
\Gamma_{\mu} (p,k) = -i \gamma_{\mu} G(p^2,k^2).
\ee
The full vertex {\sl ansatz} (\ref{vertex}) appears only once in $V_2$
to avoid double-counting.  This treatment of the vertex is consistent
with that in Dyson--Schwinger calculations; when we turn to the
effective potential calculation, the function $G$ will be set to
unity.  In order to take account of higher loop effects in the vertex,
one has to consider the ``trilocal'' effective action discussed
briefly above.

When these functions are substituted into the Dyson--Schwinger
equations (\ref{dss}, \ref{dsd}) the following gap equations for $A$
and $B$ can be derived:
\ba \label{dsgaps}
A(q)&=&1-\frac{\alpha}{\pi^2 N}\frac{1}{q^3} \int_0^{\infty} dk \,
\frac{kA(k)G(q^2,k^2)}{{\cal K}(k)} {\cal I}_1(k,q;\alpha),
\nonumber \\
B(q)&=&\frac{\alpha}{\pi^2 N} \frac{1}{q} \int_0^{\infty} dk \,
\frac{kB(k)G(q^2,k^2)}{{\cal K}(k)} {\cal I}_2(k,q;\alpha);
\ea
with the integrals given by:
\ba \label{angints}
{\cal I}_1 (k,q;\alpha) &=& \alpha^2 \ln \left[
\frac{k+q+\alpha}{|k-q|+\alpha} \right] - \alpha \left( k+q -|k-q|
\right) \nonumber \\
&-& \frac{1}{\alpha} |k^2-q^2| \left( k+q - |k-q| \right) + 2kq \nonumber \\
&-& \frac{1}{\alpha^2} \left( k^2 - q^2 \right) \left\{ \ln \left[
\frac{k+q+\alpha}{|k-q|+\alpha} \right] - \ln \left[ \frac{k+q}{k-q}
\right] \right\} \nonumber ; \\
{\cal I}_2 (k,q;\alpha) &=& 4 \ln \left[
\frac{k+q+\alpha}{|k-q|+\alpha} \right],
\ea
and
\be
{\cal K}(k)= A^2(k) k^2 + B^2 (k).
\ee
These equations can now be solved (using analytical and numerical
methods) for $A$ and $B$ to yield the full fermion propagator which
minimizes the effective potential.

The approach taken in this paper will differ from the conventional
Dyson--Schwinger method outlined above.  We shall use the {\sl ansatz}
(\ref{anz}) directly in the effective potential, equation
(\ref{effpot}),and demand that $V[A,B]$ be stationary with respect to
(functional) variations in $A$ and $B$.  We can then derive the ``gap
equations'' for $A$ and $B$ and compare them with the equations found
in the Dyson--Schwinger approach, equations (\ref{dsgaps}).

\section{Gap Equations from the Effective Potential.} \label{effgaps}

The effective potential comprises three natural parts; the trace over
the fermion propagator terms, the trace over the gauge boson
propagator terms, and the sum of two-particle irreducible vacuum
graphs.  The second of these will play no part in the variation to be
described: we will again use the leading order (in $1/N$) truncation
for the gauge propagator (\ref{gauge}).  After substituting
(\ref{anz}), the variation of the pure fermion part is given by:
\ba
\frac{\delta V_f [A,B]}{\delta A(q)} &=& \frac{4Ni}{(2\pi)^3} \frac{A^2 (q)
q^4 (A(q) -1) + B^2 (q) q^2 (A(q) +1)}{{\cal K}^2(q)}, \nonumber \\
\frac{\delta V_f [A,B]}{\delta B(q)} &=& \frac{4Ni}{(2\pi)^3}
\frac{A(q)B(q)q^2(A(q)-2) + B^3(q)}{{\cal K}^2(q)}.
\ea

When these variations alone are set to vanish, the equations have the solutions
$A=1, B=0$, consistent with the free propagators.  The variation of
the loop part is:

\ba
\frac{\delta V_2 [A,B]}{\delta A(q)} = &-& \frac{ie^2 N}{2\pi^3}
\frac{2A(q)B(q)q^2}{{\cal K}^2 (q)} \int_k \frac{B(k)}{{\cal K}(k)}
\frac{2}{(k-q)^2} \frac{1}{1+\alpha / |k-q|}\nonumber \\
&-& \frac{ie^2 N}{2 \pi^3} \left[\frac{2A^2 (q)q^2}{{\cal K}^2 (q)} -
\frac{1}{{\cal K}(q)} \right] \int_k \frac{A(k)}{{\cal K}(k)} \frac{
2}{(k-q)^4} \frac{q\cdot(k-q) k \cdot (k-q)}{1+ \alpha /
|k-q|}, \nonumber \\
\frac{\delta V_2 [A,B]}{\delta B(q)} = &-& \frac{ie^2 N}{2\pi^3}
\left[\frac{2B^2 (q)}{{\cal K}^2 (q)} - \frac{1}{{\cal K}(q)} \right]
\int_k \frac{B(k)}{{\cal K}(k)}
\frac{2}{(k-q)^2}\frac{1}{1+\alpha / |k-q|} \nonumber \\
&-& \frac{ie^2 N}{2 \pi^3} \frac{2A(q)B(q)}{{\cal K}^2 (q)} \int_k
\frac{A(k)}{{\cal K}(k)} \frac{ 2}{(k-q)^4} \frac{q\cdot(k-q) k
\cdot (k-q)}{1+\alpha /|k-q|}.
\ea

Computing the resulting angular integrations, performing some
simplifications, and setting the variations to vanish we obtain the
``raw'' gap equations:

\ba \label{gap1}
0 &=& A^2 (q) q^4 (A(q)-1) + B^2 (q) q^2 (A(q)+1) 
- \frac{2 \alpha}{N\pi^2} 2 A(q)B(q) q^2 \! \int \! dk \, {\cal J}_1
(k,q;\alpha) \nonumber \\
&-& \frac{2 \alpha}{N\pi^2} \left( A^2 (q)q^2 - B^2 (q)\right) \int
dk \, {\cal J}_2 (k,q;\alpha), \\
0 &=& A(q)B(q)q^2(A(q)-2) + B^3(q)
+ \frac{2 \alpha}{N\pi^2} \left( A^2(q)q^2 - B^2(q) \right) \int dk \,
{\cal J}_1 (k,q;\alpha) \nonumber \\
&-& \frac{2 \alpha}{N\pi^2} 2A(q)B(q) \int dk \, {\cal J}_2
(k,q;\alpha)\label{gap2};
\ea
where the integrals are given by:
\ba
{\cal J}_1 (k,q,\alpha) &=& k^2 \frac{B(k)}{{\cal K}(k)} \frac{2}{kq} \ln \left[
\frac{k+q+\alpha}{|k-q|+\alpha} \right], \nonumber \\
{\cal J}_2 (k,q;\alpha) &=& 2 k^2 \frac{A(k)}{{\cal K}(k)}
\left( \, \frac{(k^2-q^2)^2 - \alpha^4}{4kq\alpha^2} \ln \left[
\frac{k+q+\alpha}{|k-q|+\alpha} \right]\right.\nonumber \\
&-& \left. \frac{(k^2 - q^2)^2 }{4kq\alpha^2} \ln
\left[\frac{k+q}{|k-q|} \right]
+ \frac{\alpha}{4kq} \left[ k+q - |k-q| \right] -
\frac{1}{2} \right. \nonumber \\
&+& \left.\frac{(k^2-q^2)^2}{4kq\alpha} \left[ \frac{1}{|k-q|} -
\frac{1}{k+q} \right] \right).
\ea

Since the momentum integrals appearing in equations (\ref{gap1},
\ref{gap2}) are the same we can eliminate each of them in turn: after
a little simplification this yields the usual gap equations obtained
from the Dyson--Schwinger approach, equations (\ref{dsgaps}), up to
the replacement $G \rightarrow 1$ which is imposed by considering
only a ``bilocal'' effective action.

\section{Concluding Remarks.}

The fact that the composite operator effective potential yields the
same gap equations (\ref{dsgaps}) whether the {\sl ans\"atze} are put
in before or after the functional variation demonstrates explicitly
that it is equivalent to the Dyson--Schwinger method for QED$_3$.  It
is by no means obvious from the ``raw'' gap equations (\ref{gap1},
\ref{gap2}) that they will reduce on simplification to the normal
Dyson--Schwinger results; since the effective potential and
Dyson--Schwinger methods are frequently applied to QED$_3$ it is
important to have checked their equivalence within the framework of
the approximations normally used.  The consistency here is a basis for
confidence that the {\sl ans\"atze} used are sufficiently consistent
and general: in other work \cite{gac} it has been found that overly
restrictive {\sl ans\"atze} lead to differences between
Dyson--Schwinger type approaches and the results of composite operator
effective action computations.

To better model high-T$_{\rm c}$ superconductivity, finite temperature
effects should really be included.  The conclusion that can be drawn
from the present calculation is that using the composite operator
effective potential for such work (see e.g. \cite{gpi}) is not
likely to yield additional information nor to simplify the computation
compared with a finite temperature Dyson--Schwinger method.

At both finite and zero temperature the choice of vertex function
remains the major problem with this sort of calculation: how to make
an {\sl ansatz} that is consistent with the Ward--Takahashi identity,
when that identity requires the full fermion propagator.  This problem
is not resolved by using the ``bilocal'' composite operator effective
potential; however, in principle this problem could be solved
completely by using a ``trilocal'' composite operator effective action,
leading to a {\sl prediction} for the vertex.  However the calculation
would be extremely difficult.  The most promising way to attack this
particular problem is by using the non-local gauge \cite{amm}, in
which the wave-function renormalization $A$ is forced to be unity.
This approach does not work well within the composite operator
effective potential formalism, for the resulting analogue of equation
(\ref{gap2}) for $B$ is intractable.

\subsection*{Acknowledgments.}

The author would like to thank G. Amelino--Camelia for suggesting the
problem, and to thank I Aitchison, G. Amelino--Camelia and
N. Mavromatos for helpful and entertaining discussions.  Thanks are
also due to P.P.A.R.C. (U.K.) for a research studentship (number
96314661).

\end{document}